\pdfoutput=1
\documentclass[aps,preprint,nofootinbib,preprintnumbers,citeautoscript]{revtex4}
\usepackage{amsfonts,amsthm,latexsym,graphicx,verbatim}
\usepackage{slashed}
\usepackage{amsmath}
\usepackage{amssymb}
\usepackage{epsfig}
\newcommand{\be}{\begin{equation}}
\newcommand{\ee}{\end{equation}}
\newcommand{\bea}{\begin{eqnarray}}
\newcommand{\eea}{\end{eqnarray}}
\def\bse{\begin{subequations}}
\def\ese{\end{subequations}}

\def\IZ{\relax\ifmmode\hbox{Z\kern-.4em Z}\else{Z\kern-.4em Z}\fi}

\newcommand{\reef}[1]{(\ref{#1})}

\def\del{{\partial}}

\newcommand{\ie}{{\it i.e.,}\ }

\newcommand{\N}{{\mathcal N}}

\newcommand{\cN}{{\mathcal N}}

\newcommand{\cD}{{\mathcal D}}

\newcommand{\hu}{{\hat u}}

\def\bi{\begin{itemize}} \def\ei{\end{itemize}}

\def\({\left(} \def\){\right)}
\def\[{\left[} \def\]{\right]}
\begin{document}
\title{ \center{Dual description of a 4d cosmology}}
\author{Michael Smolkin and Neil Turok \\[0.1cm]
\it Perimeter Institute for Theoretical Physics, \\
Waterloo, Ontario N2L 2Y5, Canada.
 }
\begin{abstract}{\noindent M-theory compactified on $S^7/Z_k$ allows for a four-dimensional, asymptotically AdS cosmology. The holographic dual consists of ABJM theory with a non-supersymmetric marginal deformation. At weak 't~Hooft coupling the dual theory possesses a UV fixed point, allowing the construction of a conformal field theory dual. The dual theory has an infinite family of instantons, which we are able to sum completely. The resulting theory spontaneously breaks global conformal symmetry but is manifestly unitary and non-singular. Our findings point to a holographic resolution of the singularity, measure and conformal factor problems in quantum cosmology. }
\end{abstract}

\maketitle

\section{Introduction}

Further progress in our description of the early universe requires the resolution of three fundamental and related problems. The first and most basic is the big bang singularity at which general relativity breaks down. The second is that the canonical (phase space) measure for cosmological backgrounds is unbounded, so that without some further constraint there is no well-defined statistical ensemble. The third, known as the conformal factor problem, is that the Euclidean action for general relativity is unbounded below. This compromises any efforts to employ a path integral description capable of covariantly incorporating the full symmetries of the theory.

Fortunately, a powerful new tool for quantum gravity has recently emerged, namely the AdS/CFT correspondence~\cite{Maldacena:1997re}. In this Letter we show how, within a certain compactified version of M-theory, AdS/CFT can be extended to obtain a well-defined dual description of a singular 4d cosmology. We explicitly construct a Euclidean path integral measure for the dual CFT, making strong use of a UV fixed point where Weyl-invariance becomes exact.  The cosmologies we study are not realistic: they possess a negative cosmological constant and negative space curvature. And the dual theory is only analytically tractable at weak gauge coupling, which corresponds to the bulk AdS radius being shorter than the string length. Nevertheless, we believe the methods developed here constitute a significant step towards tackling the abovementioned fundamental puzzles. 

We work in one of the simplest and best-defined theoretical frameworks for 4d cosmology, namely M-theory compactified on $S^7/Z_k$. With supersymmetric boundary conditions, the bulk background spacetime is anti-de Sitter (AdS) spacetime. The AdS/CFT dual was identified by Aharony {\it et al.} \cite{Aharony:2008ug}, as a supersymmetric Chern-Simons gauge theory known as ABJM theory.  Non-supersymmetric but still asymptotically AdS  boundary conditions allow for more interesting big crunch cosmologies in the bulk~\cite{Hertog:2004dr,Hertog:2004rz,Hertog:2005hu}. They consist of a time-symmetric  $O(3,1)$-invariant bubble solution within an asymptotically-AdS spacetime. The bubble contains an infinite open FRW universe collapsing to a big crunch. Outside the bubble, the natural $O(3,1)$-invariant boundary is  3d de Sitter spacetime $dS^3$. The boundary avoids both past and future singularities in the bulk and hence one can hope for a nonsingular holographic description. 

The cosmological dual is a marginal deformation of ABJM theory, as described in Ref.~\cite{Craps:2009qc}. It is analytically tractable for large $k$, {\it i.e.}, at weak gauge coupling. An important observation was made in Ref.~\cite{Craps:2009qc}: at large $k$, the  deformation coupling has an UV fixed point, meaning that the dual theory is perturbatively ultraviolet-complete and hence, in principle, provides a complete description of the cosmological singularity in the bulk.  Heuristically, as the bulk geometry shrinks to zero at the crunch, the propagating degrees of freedom migrate outward to the boundary, where they are described by the UV limit of the boundary theory in accordance with the UV-IR correspondence. 

\begin{figure}
\begin{center}
\includegraphics[width=7cm,height=7cm]{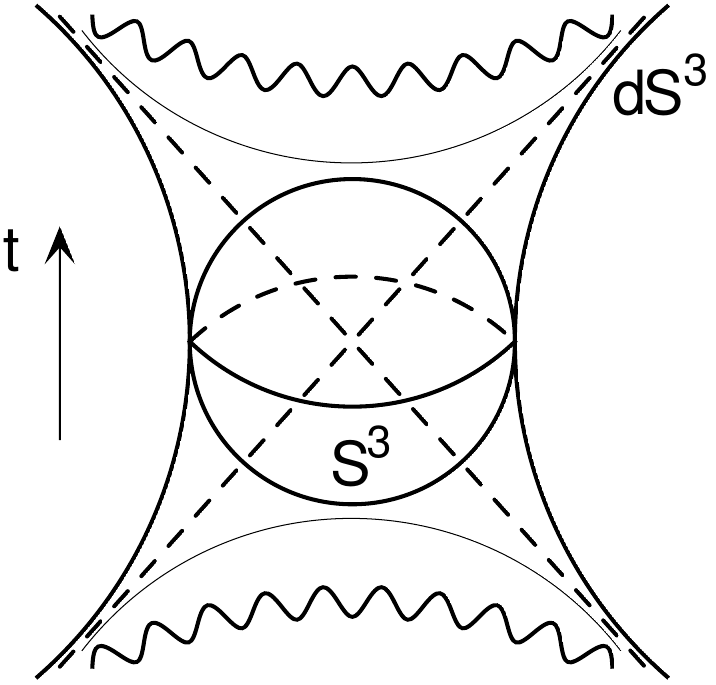}
\caption{An asymptotically-$AdS^4$ cosmology with a $dS^3$ holographic boundary. The bulk contains a time-symmetric bubble which expands to the future and the past, in each case containing an infinite open FRW universe which collapses to a big crunch singularity. We define the dual theory by analytic continuation from $S^3$.}
\label{cosmofig}
\end{center}
\end{figure}

Our main tool for the study of the dual theory is the large $N$ expansion. As Ref.~\cite{Craps:2009qc} showed, the ABJM theory at large $k$ reduces to an $O(N)$ vector model in three dimensions. The marginal deformation is governed by a sextic coupling\footnote{Actually, away from $g_6=g_6^*$, there are several sextic couplings, but we focus on the one which dominates at the UV fixed point: see~\cite{Craps:2009qc}.} $g_6$, which possesses an ultraviolet fixed point at  $g_6=g_6^*=192$~\cite{Pisarski}, up to $1/N$ corrections. However, there has been a longstanding puzzle about this theory since the work of Bardeen {\it et al.}~\cite{Bardeen:1983rv}, who showed that for $g_6>g_6^c=16 \pi^2$, then for conventional quantization in flat spacetime, the quantum Hamiltonian is unbounded below. Since one can always consider high energy processes (or vacuum fluctuations) which probe the UV and in which $g_6$ runs up to $g_6^*>g_c$, it would seem that in flat space the theory is unstable.  

Motivated by AdS/CFT cosmology,  we have reconsidered the model for $g_6>g_6^c$, but instead defined on an $S^3$ (or its Lorentzian continuation $dS^3$). We can consider the theory precisely at $g_6=g_6^*$, for which it is exactly Weyl-invariant. 
$S^3$ (or $dS^3$) is the most natural boundary for asymptotically $H^4$ (or $AdS^4$) metrics such as we consider here. As usual at large $N$, we define the theory by introducing auxiliary fields. For $g_6>g_c$, we shall show that suitable integration contours for these fields may be chosen in the complex plane so that the Euclidean path integral converges and is real.  Our procedure is similar in spirit to the Gibbons-Hawking-Perry proposal for Euclidean quantum gravity~\cite{Gibbons:1978ac} but of course much simpler to implement in the dual theory than it is in gravity. For $g_c<g_6<g_6^*$, and at finite $N$, $g_6$ decreases in the IR but we can consider the theory on a small enough $S^3$ that $g_6$ never flows below $g_c$. For $g_6>g_6^*$, $g_6$ increases in the IR and we can consider the theory on an arbitrarily large $S^3$.

One of the subtleties of the dual theory at its UV fixed point is that, due to its scale-invariance, it allows an infinite family of instantons. These correspond to bulk instantons responsible for nucleating a cosmological bubble as illustrated in Fig. 1. Any such instanton breaks the AdS isometry group $O(3,2)$ (or its Euclidean version $O(4,1)$) down to $O(3,1)$ (or $O(4)$), {\it i.e.}, to cosmological symmetry. The four broken generators act to translate the bubble around in the 4d bulk. Since bubbles can nucleate anywhere in the infinite bulk, one might expect an infinite instability rate~\cite{Harlow:2010az}. 

The dual CFT defined on $S^3$ (and its Lorentzian continuation $dS^3$) possesses a corresponding infinite family of instantons. Each breaks $O(4,1)$ global conformal symmetry down to the isometry group $O(4)$.  At first sight, these instantons might also appear to mediate an infinite instability. However, when we carefully define the theory so that the Euclidean path integral converges, and sum over all instantons in the Euclidean region, there is no instability.  We find  $O(4,1)$ is spontaneously down to $O(4)$ but translation symmetry on $S^3$ (or its Lorentzian continuation $dS^3$) is unbroken, and the vacuum is now manifestly stable, unitary and finite. In principle, our construction therefore provides a complete holographic description of the cosmological singularity. We explicitly calculate the boundary two-point function, show it satisfies reflection positivity, and display its short and large distance limits.

\section{The dual CFT at weak gauge coupling}

Our construction of the dual CFT begins with the $O(N)$ vector model on $S^3$, defined by the Euclidean action
\be
\mathcal{S}_E = \int_{S^3} \left[ \frac{1}{2} (\partial \vec\phi)^2 
 +{R \over 16}\vec\phi^{\,2} 
 +\frac{g_{6}}{6 N^2}(\vec\phi ^{\,2})^{3}\right].
 \ee
We re-express the interaction using a Lagrange multiplier field $s$ and an auxiliary field $\rho$:
 \be 
 \int_{S^3}\left[\frac{1}{2} (\partial \vec\phi)^2 
 +{R \over 16}\vec\phi^{\,2} +{1\over 2} s(\vec\phi^{2}-N \rho)
 +N \frac{g_{6}}{6}\rho^{3}\right],
 \ee
and then integrate out $\vec{\phi}$ in the generating functional
\be
 Z_{\vec{J}}  =\int \cD s \cD\rho
 e^{ -\frac{N}{2} \int(\frac{g_{6}}{3}\rho ^{3}-s \rho) -\frac{N}{2}  \text{Tr} \ln \hat{O}_{s}+\frac{1}{2} \langle\vec{J}, \hat O^{-1}_{s} \vec{J}\rangle}
 \, ,
 \label{part}
 \ee
 where $\hat{O}_{s}\equiv -\square + \frac{R}{ 8} +s$ and $\langle,\rangle$ is the $L^2$ norm. In this representation of $Z_{\vec{J}}$, the auxiliary field $s$ encodes  the full dynamics of the original $N$ physical degrees of freedom\footnote{Note that we study the $O(N)$ symmetric phase of the model.}, while $\rho$ has trivial dynamics since it appears algebraically in the action. Both $s$ and $\rho$ are singlets which significantly simplifies the $1/N$ expansion. The contours of integration for $s$ and $\rho$ are chosen to ensure the path integral converges. We take the theory to be defined on an $S^3$ of radius $r_0$ (so $R=6/r_0^2$). However, since there is no conformal anomaly in 3d, at the UV fixed point the theory is Weyl-invariant, and it shall prove very convenient to Weyl-transform to spheres of different radius or to $R^3$. 
  
To solve the theory, we look for a large $N$  saddle point. The resulting gap equations read
 \be
\rho= \langle x| \hat{O}_s^{\,-1} |x\rangle,	\qquad s=g_6 \rho^2.
\label{gapeq}
 \ee
The saddle point equations are Weyl invariant for any value of $g_6$ because all anomalous dimensions vanish to leading order in $1/N$~\cite{Pisarski}. Let us assume the solution is homogeneous when taken on a sphere $S^3$ of radius $a$, so we write $\rho=\bar{\rho}_a,$  $s=\bar{s}_a$. In this case the inverse of  $\hat{O}_{\bar{s}}$ coincides with the Green's function for a free massive scalar field. The solution to the gap equations is\footnote{To cancel the power law divergences resulting from the short distance expansion of $\hat{O}_{\bar{s}}^{-1}$, we introduce counterterms which are quadratic and quartic in the field $\vec \phi$  and tune the associated renormalized couplings to zero, \ie we set the theory at the tricritical point.}
 \be
\bar{\rho}_a a =-{\mathcal{N} \coth\( \pi\mathcal{N} \) \over 4 \pi}, \quad \bar{s}_a a^2 =\cN^2+{1\over 4} =g_6 \bar{\rho}^2_a a^2, 
 \label{gap-sphere}
 \ee
where we define $\cN\equiv \sqrt{\bar s_a a^2 - {1\over 4}}$. For all $g_6>g_c=16\pi^2$, there is a positive solution $\bar{s}_a=C(g_6) a^{-2}$, with $\infty>C(g_6)>0$ as $g_c<g_6<\infty$ and $C(g_6^*)\approx1.38$.  For any $a$, the solution may be Weyl transformed to a sphere of radius $r_0$.  Transforming each solution stereographically first to the plane $R^3$ and then to a sphere of radius $r_0$, one obtains the following infinite family of instantons on  $S^3$ of radius $r_0$, 
\be
\bar s (\hat\eta\,;\,a,\hat u)={4\, \bar{s}_a \over \big[1+\hat\eta\cdot\hat u+{r_0^2\over a^2}(1-\hat\eta\cdot\hat u)\big]^2}~,
 \label{inst}
 \ee
where  $\hat\eta $ is any point on $S^3$, $\hu$ is the instanton's centre and $a$ it's size. For all $a\neq r_0$, each instanton, if individually continued to Lorentzian time, {\it i.e.}, to $dS^3$, becomes singular. However, as we shall see, if we {\it first} sum over all of these instantons in the Euclidean region $S^3$ and {\it only then} continue to Lorentzian time, the theory is completely regular on $dS^3$.

\section{Generating functional in the instanton background}

\begin{figure}
\begin{center}
\includegraphics[width=10cm,height=10cm]{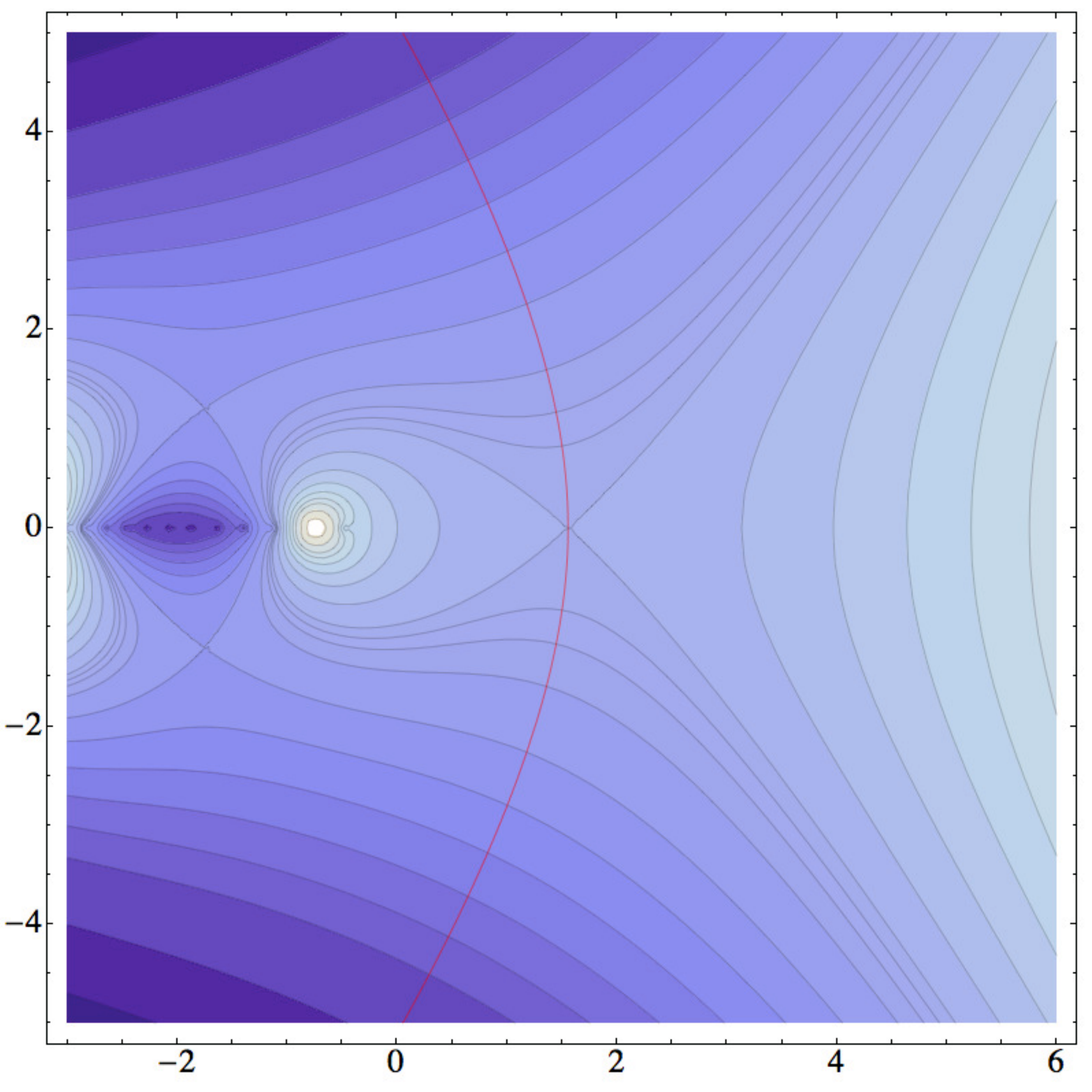}
\caption{A contour plot of the logarithm of the magnitude of the integrand in Eqn. (\ref{parta}), using (\ref{flucdet}), for the homogeneous mode of the auxiliary field $\bar{s}$, plotted in the $\bar{s}$-plane. Light is positive, dark negative. The saddle point at positive real $\bar{s}$ and the steepest descent contour used to define the theory are shown. Notice the singularity at $\bar{s}=-{3\over 4}$, where (\ref{flucdet}) diverges, causing the integrand to diverge as $(\bar{s}+3/4)^{-\pi N/2}$. This singularity prevents the chosen contour from being deformed to lower action contours in the complex $\bar{s}-$plane. Parameters used in this plot were $N=15, g_6=g_6^*=192$.}
\label{vertex6}
\end{center}
\end{figure}

To define the path integral about each instanton, we must find a contour in the space of auxiliary fields $s$ and $\rho$ for which the path integral converges. For each instanton we work in the Weyl frame in which it is homogeneous on the $S^3$. In this frame, it is straightforward to see that the operator associated with quadratic fluctuations around a given instanton has one negative mode - the homogeneous mode, $\bar{s}$, and four zero modes - the $l=1$ harmonics, which correspond to translations and dilatations on $R^3$. All higher harmonics have positive eigenvalues. We determine the integration contour for the homogeneous mode by steepest descent. Since the higher harmonics occur at least quadratically in the action, for this purpose they may consistently be set to zero. We proceed by integrating out the homogeneous mode $\bar{\rho}$ to obtain for the homogeneous mode $\bar{s}$ the following contribution to $Z_0$:
 \be
 Z_0 \sim -i \int  d\bar{s}  {\rm Ai}(\alpha \bar{s})
 e^{-\frac{N}{2}  \text{Tr} \ln \hat{O}_{\bar{s}} },
 \label{parta}
 \ee
where ${\rm Ai}$ is the Airy function, $\alpha=\(U N / 2\)^{2\over 3} g_6^{-{1\over 3}}$ and $U$ is the volume of $S^3$. The fluctuation determinant becomes
\be
-{N\over 2} \text{Tr} \ln \hat{O}_{\bar{s}}={UN \over 8\pi} \int^{\bar s} d\bar{s} \, \mathcal{N} \coth\( \pi \mathcal{N} \),
\label{flucdet}
\ee
plus an unimportant constant. For positive real $\bar{s}$ and large $N$ we have  ${\rm Ai} (\alpha \bar{s}) \sim (\alpha \bar{s})^{-{1\over 4}} e^{-{U N\over 3 \sqrt{g_6} } \bar{s}^{3\over 2}}$. In (\ref{parta}) the exponents combine to produce the large $N$ saddle point solution given above. Fig.~\ref{vertex6} presents a contour plot of the logarithm of the absolute value of the integrand in the complex $\bar{s}$-plane. We also show the steepest descent contour, which runs through the saddle point at the centre of the figure and asymptotes, at large $|s|$, to $\theta = \pm 2\pi/3$.  By the symmetry of the contour under reflections in the real $\bar{s}$-axis, the reality of the functions in \reef{parta} at real $\bar{s}$ and the Schwarz reflection principle, the integral over $\bar{s}$ in \reef{parta} is real. A similar argument applied to both the $\bar{\rho}$ contour which we implicitly used in obtaining the Airy function, and the $\bar{s}$ contour just defined, shows that the full generating functional $Z_{\vec{J}} $ for $\vec{\phi}$ correlators is real, to all orders in $1/N$.

The leading, saddle point approximation is then:
 \be
 Z_0=e^{- S_\text{inst} }\sum_{a,\hu} \int d(-i\bar{s})  \prod_{l=2}^{\infty} d s_l \,  e^{-S_{\text{fluc}}(a,\hu;\bar{s};s_l)}~,
 \label{instZ}
 \ee
where $S_\text{inst}\propto N $ is the instanton action. The integral over the negative mode $\bar{s}$ converges on the steepest descent contour. The four zero modes correspond to the $a$ and $\hu$ moduli fixing the instanton's size and location on the $S^3$ of radius $r_0$. Since $\bar{s}_a^2\sim a^{-2}$ it follows from Eq. \reef{inst} that the instanton is invariant under inversion $\hu\rightarrow -\hu$ accompanied by rescaling $a\rightarrow r_0^2/a$. To avoid double counting, we must therefore restrict the summation over $a$ to $r_0\leq a < \infty$. For all such instantons, we can regularize the theory by Weyl transforming to flat space and introducing a flat-space UV cutoff in the field theory satisfying $\Lambda^{-1}\ll a$. Hence,
\be
 \sum_{a,\hu}\equiv\int_{r_0}^\infty {da \over a}\int_{S^3} C_d \, C_t^3~,
\ee
where $C_d$ and $C_t$ are the norms of the dilation and translation modes respectively, computed using the moduli space metric (see e.g. \cite{Tong}):
 \be
 C_d\sim\sqrt{ r_0^3 a(a^2-a r_0+r_0^2)\over (a+r_0)^4}~, \quad
 C_t\sim\bigg|{a-r_0\over a+r_0}\bigg|.
 \label{norms}
 \ee
Note that $C_d$ diverges as $r_0\to\infty$. This indicates that zero mode associated with dilation is not normalizable in flat space. This divergence is avoided on $S^3$.

As mentioned, at the fixed point we expect the model to exhibit global conformal symmetry $O(4,1)$. However, each instanton spontaneously breaks $O(4,1)$ to $O(4)$. This is seen by Weyl transforming the vev for $s$ (which has scaling dimension $\Delta=2$) to $R^3$, on which
 \be
 \langle s\rangle\big|_{R^3}= \langle s\rangle\big|_{S^3} \bigl({2a^2\over a^2+r^2}\bigr)^2~.
 \label{vev}
 \ee 
This clearly breaks dilation symmetry generated by $D=-i(x^\nu\del_\nu+\Delta)$, special conformal symmetry generated by $K_\mu=2x_\mu D+ix^2\del_\mu$ and translations generated by $P_\mu=-i\del_\mu$. It is invariant under the $O(4)$ symmetry generated by  $L_{\mu\nu}=i(x_\mu\del_\nu-x_\nu\del_\mu)$ and $L_{\mu 4}=K_\mu+a^2 P_\mu$, corresponding to rotations on the $S^3$. Therefore, as claimed, each instanton breaks $O(4,1)$ to $O(4)$. However, because translation invariance on $R^3$ is broken, the four broken generators do not lead to a dilaton or Goldstone modes. Instead, they generate the instanton moduli space, which we integrate over in the path integral. 

Through a straightforward computation one can evaluate the Gaussian integral in \reef{instZ}:
 \be
 \int  \prod_{l=2}^{\infty} d s_l \,  e^{-S_{\text{fluc}}(a,\hu;\bar{s};s_l)}
={(a\Lambda)^\kappa\over a^4}+\ldots , 
 \label{gaussint}
 \ee
where $\Lambda$ is a UV cutoff, ellipses denote either finite terms or scheme dependent divergences which will not be important for our further discussion, and
 \be
 \kappa= -{\big(\N\coth(\pi\N)\big)^3\over 2(4\pi)^3}\,g_6^2\bigg(1-{g_6\over 192}\bigg)~.
 \ee
As a result, we obtain
\be
 Z_0 \sim  (r_0 \Lambda)^\kappa e^{-S_{\text{inst}} } ~.
\ee
One can see that $Z_0$ is cutoff independent provided $g_6$ satisfies the RG flow given in \cite{Pisarski,Craps:2009qc}.

\section{two-point correlator}

Taking the derivatives of the generating functional \reef{part} with respect to $\vec J$ and integrating over the saddle point yields the following expression for the $2$-point correlator of the scalar field $\vec\phi$ on $S^3$ of radius $r_0$:
\begin{multline}
\langle  
 \phi_0^m(\hat \eta_1)\phi_0^n(\hat \eta_2)\rangle
 = \delta^{mn} Z_0^{-1}\sum_{a,\hu}~
 O_{\bar s}^{-1}(\hat \eta_1,\hat \eta_2)  
 \\
 \times \int d(-i\bar{s})  \prod_{l=2}^{\infty} d s_l \,  e^{-S_{\text{fluc}}(a,\hu;\bar{s};s_l)}
 +\mathcal{O}(1/N)~,
 \label{correlators}
 \end{multline}
where $\hat \eta_i$ are angular coordinates on the $S^3$. 

Weyl invariance implies a simple relation between $O_{\bar s}^{-1}(\hat \eta_1,\hat \eta_2)$ on $S^3$ of radius $r_0$ and the appropriate massive scalar field propagator on $S^3$ of  radius $a$. This yields
 \be
  O_{\bar s}^{-1}(\hat \eta_1,\hat \eta_2)=-{\sinh\[\N(\chi_a-\pi)\] \over 4\sqrt{2}\,\pi r_0 \, \sinh (\pi\N )\, \cos{\chi_a\over2}(1-\hat\eta_1\cdot\hat\eta_2)^{1/2}}
  ~,
  \label{proptrans2}
 \ee
where $\chi_a$ is the angle between the stereographic projections of the points $\hat \eta_1$ and $\hat \eta_2$ onto a sphere of radius $a$
 \be
 \sin^2{\chi_a\over 2}={2 \, a^2 r_0^2(1-\hat\eta_1\cdot\hat\eta_2) \over \prod_{i=1,2}\big[a^2(1-\hat\eta_i\cdot\hat u)+r_0^2(1+\hat\eta_i\cdot\hat u)\big]}~.
 \label{proptrans3}
 \ee
As a result, the short distance expansion of the  $\vec\phi$ propagator can be readily evaluated
 \be
\langle 
 \phi_0^m(\hat \eta_1)\phi_0^n(\hat \eta_2)\rangle
 = \delta^{mn} \bigg({1\over 4\pi r_0\chi_{r_0}}  - {f(g_6)\cN \coth(\pi\cN)\over \pi r_0}
  +\ldots\bigg)
 \ee
where $f(g_6)$ is a known positive function, with $f(g_6^*)\approx 0.2$, compared with the equivalent term for a free massive scalar with mass $r_0^{-1}\sqrt{\cN^2+{1\over 4}}$, where  $f(g_6)$ takes the value ${1\over 4}$. 

The first term on the right hand side represents the Euclidean propagator of the free conformally coupled massless field. To leading order in the short distance expansion any correlator is given by a product of such propagators and the theory becomes essentially free. The next to leading, constant term can be interpreted as the regularized expectation value of $\phi_0^2$.

In a unitary theory, the Euclidean two-point function should satisfy reflection positivity. In our case, this simply means it should be positive. This follows straightforwardly from the positivity of the two point function for a massive scalar on $S^3$, and all of the Weyl factors. Thus the entire integrand in  \reef{correlators}  is positive. To explore the large distance behaviour of the two-point function one needs to analytically continue it to de Sitter space. Under the analytic continuation $Z\equiv\hat\eta_1\cdot\hat\eta_2$ becomes the SO(3,1)-invariant de-Sitter dot product, and the large distance limit is tantamount to the large  $|Z|$ behaviour. Taking this limit, we get from eqs. \reef{proptrans2} and  \reef{proptrans3},
$ O_{\bar s}^{-1}(\eta_1,\eta_2)\sim Z^{-1/2}~ \Rightarrow ~\langle \phi^c(\eta_1)\phi^c(\eta_2)\rangle\sim Z^{-1/2},$ 
 the behavior expected for a conformally coupled massless scalar.

\section{Conclusions}

In summary, we have constructed the holographic dual to a 4d M-theory cosmology, for weak gauge coupling, {\it i.e.},  when the bulk is in the stringy regime. Several lines of development can now be pursued. First, through understanding the bulk behavior in the Einstein gravity regime, including perturbations, we can study the behavior of the holographic dual at strong gauge coupling. Second, through a study of the asymptotics on the $dS^3$, we can study the double holographic dual, {\it i.e.}, the 2d dual CFT defined
on the past and future $S^2$ boundaries of the $dS^3$. This may in turn allow us to propagate information across the cosmological singularities in a unique manner. Further details, and comparisons to other work~\cite{Maldacena:2010un}, will be presented elsewhere. 

\vfill\eject

{\bf Acknowledgements:} It is a pleasure to thank T.~Banks, L.~Boyle, B.~Craps, D.~Harlow, S.W.~Hawking, T.~Hertog, Z.~Komargodski, R.C.~Myers, S.~Shenker  and L.~Susskind for many helpful comments. Research at Perimeter Institute is supported by the Government of Canada through Industry Canada and by the Province of Ontario through the Ministry of Economic Development and Innovation.

\end{document}